\documentclass[
	letterpaper,
	10pt, 
        unnumberedsections,
	twoside 
]{styles}

\runninghead{Optimizing Closed Payment Networks on the Lightning Network: Dual Central Node Approach} 

\footertext{\textit{}} 

\title{Optimizing Closed Payment Networks on the \\ Lightning Network: Dual Central Node Approach} 

\author{%
	Jeffy Yu\\
}

\date{\footnotesize Parallel Polis\\ San Francisco State University\\}

\renewcommand{\maketitlehookd}{%
	\begin{abstract}
		\noindent The Lightning Network, known for its millisecond settlement speeds and low transaction fees, offers a compelling alternative to traditional payment processors, which often have higher fees and longer processing times. This is particularly significant for the unbanked population, which lacks access to standard financial services. Our research targets businesses looking to shift their client-to-client payment processes, such as B2B invoicing, remittances, and cross-border transactions, to the Lightning Network. We compare the efficiency of interconnected mesh nodes (complete graph topology) with central routing nodes (star graph topology), with a specific focus on the 'dual central node' approach. This approach introduces features like circular rebalancing, redundancy, and a closed network system. Through a basic SimPy model, we assess the network's throughput in a 100-node scenario. While this approach centralizes a technology initially designed for decentralization, it fosters broader enterprise adoption of Bitcoin-based payment networks and encourages participation in the decentralized financial ecosystem. Our study also considers the regulatory implications of using central routing nodes, possibly classified as payment processors under Money Transmission Laws (MTL). These findings aim to contribute to the discourse on the Lightning Network’s application in business, highlighting its potential to drive shifts in financial technology towards more decentralized systems.
	\end{abstract}
}

\begin{document}

\maketitle 


\section{1. Introduction}

In the rapidly evolving landscape of financial technology, the quest for efficient, secure, and inclusive payment systems has led to significant innovations. Among these, Bitcoin emerged as a groundbreaking digital currency, leveraging blockchain technology to facilitate peer-to-peer transactions without the need for traditional financial intermediaries [1]. Despite its transformative potential, Bitcoin's scalability challenges, manifested in the form of increased transaction fees and processing times [2], have prompted the development of the Lightning Network—a second-layer protocol designed to enable instant and cost-effective payments [3].

The Lightning Network operates by establishing payment channels between users, which allow for a multitude of transactions to occur off the main blockchain, thus alleviating congestion and reducing fees compared to mainnet [4]. However, the network's topology has been a subject of scrutiny, with concerns regarding routing failures and the equitable distribution of funds across channels [5, 6]. In the beta stage of the network, the probability of a payment success using a few dollars was 70\%, while the success rate for a payment of about \$200 was 1\% [5]. Since 2020, the network has demonstrated reduced efficiency for amounts exceeding 0.0001 BTC, with the success rate for routing a payment of 0.05 BTC dropping from 37.29\% in 2018 to just 13.61\% in 2022 [6]. Other research finds the probability of successfully routing a transaction through the Lightning Network is 100\% for amounts below 0.001 BTC, after that threshold said chances decrease dramatically reaching zero at less than 0.1 BTC [7]. While there are benefits to speed and fees, the level of transaction unreliability makes the network suboptimal for enterprise payment systems.

To address these issues, our research delves into the optimization of payment networks built on the Lightning Network, particularly for businesses seeking to modernize their payment infrastructure for activities such as B2B invoicing, remittances, cross-border transactions, and other purposes.

Our research centers on a comparative analysis of two principal network topologies applied to the Lightning Network: the interconnected mesh topology, typified by a complete graph where nodes are fully interconnected, and the centralized hub-and-spoke topology, resembling a star graph with a central routing node at its core, facilitating all transactions. We introduce an innovative 'dual central node' approach, which integrates circular rebalancing and redundancy protocols into a closed network system to bolster throughput and dependability.

The dual central node framework features a primary central node dedicated to transaction routing, alongside a secondary, dormant node that provides backup support and facilitates the rebalancing of the network's liquidity. This design enables linear scaling of channel requirements, diverging from the quadratic channel growth inherent to fully meshed networks. Through the deliberate manual establishment of channels and the exclusion of external connections, the network preserves its closed status, reinforcing the privacy and security for its participants.

To quantify the efficiency gains of our proposed topology, we employ mathematical modeling to compare the number of channels required in complete and star graphs. In a complete graph, the number of channels grows quadratically with the number of nodes, following the formula \( E = \frac{n(n - 1)}{2} \), where \( E \) represents the number of edges or channels, and \( n \) is the number of nodes. In contrast, a star graph exhibits linear growth in channel requirements, with \( E = n - 1 \), where the central node is directly connected to all other nodes, resulting in a more scalable and cost-effective structure.

For liquidity optimization within the star graph topology, we leverage linear programming techniques, which are the basis of network flow theories [8]. This mathematical approach enables us to determine the optimal distribution of liquidity across the network's channels, ensuring efficient transaction processing while minimizing rebalancing costs [9]. The linear programming model considers various constraints, including channel capacities and the need for liquidity reserves, to find the most cost-effective rebalancing strategy.

Furthermore, we explore optimization strategies for circular rebalancing within the dual central node system. Circular rebalancing is a process that allows for the redistribution of liquidity in a closed loop, ensuring that each channel within the network maintains optimal liquidity levels for transaction processing [10, 11, 12, 13]. By solving an optimization problem that minimizes the number of transactions and the amount of liquidity moved [14], we can execute rebalancing operations that are both efficient and economical.

Through a SimPy-based simulation [15], we evaluate the network's performance, focusing on its ability to handle a high volume of transactions with minimal downtime. The simulation provides insights into the network's long-term throughput capabilities, highlighting its potential to scale and maintain robustness over time.

In addition to the technical and operational aspects, our study also contemplates the regulatory landscape surrounding the use of central routing nodes. We consider the potential classification of these nodes as payment processors under Money Transmission Laws, exploring the implications for businesses operating within the network.

By centralizing certain elements within the decentralized framework of the Lightning Network, our dual central node approach seeks to strike a balance between efficiency and the ethos of decentralization. It aims to facilitate the broader adoption of Bitcoin-based payment networks by enterprises, thereby fostering greater participation in the decentralized financial ecosystem.

Given the minimal existing literature on closed payment network structures targeted at businesses to transition their payment systems to the Lightning Network, our findings aim to contribute to the movement towards these use cases.


\section{2. Background}

Since its creation, Bitcoin has faced limitations in its transaction processing capacity. The design of Bitcoin's protocol allows for a new block to be added to the blockchain approximately every ten minutes, capping the system at a maximum of about seven transactions per second [16]. This is starkly contrasted by the capabilities of traditional payment systems like Visa, which routinely handles around 2,000 transactions per second and can surge to accommodate several thousand during peak times [16, 17].

In the Bitcoin network, miners are responsible for creating and appending new blocks to the blockchain. During periods of high demand, these miners can command higher transaction fees. A notable instance of this occurred in 2017 when transaction fees surged from under one dollar to almost forty dollars at their peak [18]. The fee amount is primarily influenced by the backlog of transactions awaiting confirmation on the blockchain, and is largely independent of the transaction volume in terms of Bitcoin value. Consequently, while the blockchain can be exceedingly cost-effective for transferring large sums—capable of moving millions of dollars worth of Bitcoin for mere cents—it can become prohibitively expensive for everyday transactions and micropayments [19]. This discrepancy has fueled a growing interest in exploring blockchain solutions for financial applications, as they present a promising avenue for innovation in the sector [20].

The Lightning Network (LN) is a second-layer protocol that operates on top of the Bitcoin blockchain. It is designed to facilitate high-speed, low-cost transactions by enabling users to transact directly with one another without the need to broadcast every transaction to the blockchain. This is achieved through a network of payment channels, which are established between pairs of users [3]. Essential components include:

\textbf{Unspent Transaction Outputs (UTXOs):} UTXOs are the fundamental building blocks of Bitcoin transactions. When a transaction occurs on the Bitcoin blockchain, it consumes one or more UTXOs as inputs and creates new UTXOs as outputs. These outputs then become the inputs for future transactions. In the context of the Lightning Network, when a payment channel is opened, the Bitcoin committed by each party is locked into a UTXO through a funding transaction. This UTXO is held in a multi-signature wallet, which requires both parties to agree to any changes in the distribution of the funds. The channel's state, which reflects the balance of payments made between the parties, is updated off-chain with each transaction. When the channel is closed, a final settlement transaction spends the UTXO and distributes the funds on-chain according to the last agreed state of the channel [21]. 

\textbf{Lightning Channels:} A Lightning channel is a two-way payment channel between two participants, for example, Alice and Bob. To open a channel, Alice and Bob each commit a certain amount of Bitcoin, which is locked in a multi-signature wallet through a funding transaction on the Bitcoin blockchain. This transaction is recorded as a unique UTXO, which can only be spent with the agreement of both parties. Once the channel is open, Alice and Bob can make an unlimited number of instantaneous transactions off-chain, which are not recorded on the blockchain but are instead netted against their initial deposits. When they decide to close the channel, a settlement transaction is broadcasted to the blockchain, reflecting the final balance between the two [4, 22].

\begin{figure}[ht] 
	\centering
        \includegraphics[width=0.5\textwidth]{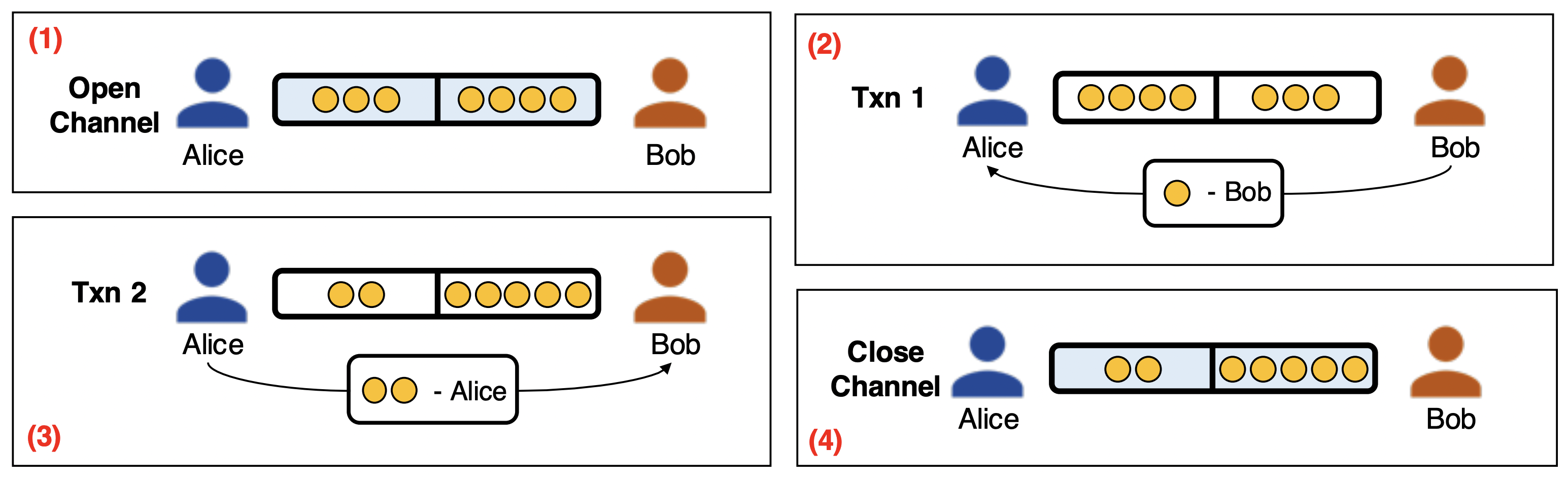}
	\captionsetup{justification=raggedright,singlelinecheck=false}\caption{Lightning Channel between Alice and Bob. Adapted from "High Throughput Cryptocurrency Routing in Payment Channel Networks" by Vibhaalakshmi Sivaraman et al., \href{https://arxiv.org/abs/1809.05088}{arXiv:1809.05088 [cs.NI]}}.
	\label{fig:tcanther}
\end{figure}

\textbf{Hashed Time-Locked Contracts (HTLCs):} The Lightning Network protocol employs advanced cryptographic techniques to facilitate secure and trustless off-chain transactions, which are essential for instant and scalable micropayments. A core component of this protocol is the HTLC, which are special types of payments conditional upon the receiver's ability to provide cryptographic proof of payment. HTLCs are time-bound, meaning the receiver must acknowledge payment within a certain timeframe, or the funds revert to the sender. This mechanism underpins the trustless nature of the Lightning Network, allowing parties to transact without the need to trust each other or a third party, while also ensuring the security and enforceability of agreements [23].

\textbf{Routing and Fees:} Transactions can be routed across multiple channels on the LN, allowing users who do not have a direct channel to still transact with one another. Each node on the network that forwards a transaction can charge a fee, which is typically composed of a base fee (a fixed amount) and a variable fee (a percentage of the transaction amount). These fees are incentives for nodes to provide liquidity and routing services to the network [22].

\textbf{Mainnet Transactions:} While the LN handles transactions off-chain, it still relies on the Bitcoin blockchain (mainnet) for opening and closing channels. These on-chain transactions require confirmations from miners and are subject to the usual transaction fees and processing times of the Bitcoin network [22].

\textbf{Nodes:} In the context of the LN, a node is an entity that participates in the network by running LN software. Nodes can open channels with other nodes, route payments, and manage their own liquidity [3, 22].

\textbf{Liquidity:} Liquidity in the LN refers to the amount of Bitcoin available in a channel that can be used to facilitate transactions. For a transaction to be successful, there must be sufficient liquidity in the right direction along the route from the sender to the receiver. When a channel is closed, the UTXO is spent, and the resulting transaction outputs reflect the final distribution of funds between the channel participants [24].

Layering a Lightning Network on top of Bitcoin enhances the system's cost, capacity, privacy, granularity, and speed, all while retaining Bitcoin's foundational trustless principle. Payments on the Lightning Network are private, as transaction details are not publicly broadcast on the blockchain [25]. This privacy extends to the route participants, who cannot identify the original sender or final recipient. The network's capacity is virtually unlimited, constrained only by the capacity and speed of individual nodes. Transactions are settled in milliseconds, and payments can be made in very small increments, down to the "dust" limit or potentially even smaller subsatoshi amounts [25]. Costs are also minimized, as the fees associated with Lightning transactions are a fraction of those on the main blockchain [26]. All transactions are conducted without the need for trusted intermediaries, preserving the decentralized ethos of the Bitcoin network while significantly expanding its capabilities. These attributes make the Lightning Network a powerful tool for enhancing Bitcoin's utility as a daily payment method and increasing the currency's fungibility by making surveillance and blacklisting significantly more challenging.

The economic incentives of the Lightning Network are designed to encourage participation and liquidity provision. Nodes earn fees for routing payments, which incentivizes them to maintain well-funded channels and reliable services. These fees are composed of a base rate plus a variable rate proportional to the payment amount. The fee structure ensures that nodes are compensated for their services without imposing excessive costs on users. The economic rationale for operating a node includes earning fees, increasing the speed and efficiency of one's own transactions, and supporting the Bitcoin ecosystem [27].

\subsection{2.1 Current State of the Lightning Network} The Lightning Network (LN), since its inception in 2018, has demonstrated considerable growth and has gained traction as a promising solution to Bitcoin's scalability constraints.

Statistically, the network has seen a steady increase in the number of nodes, channels, and overall capacity. With over 31,000 nodes now active [28], and a network total value locked that exceed \$200 million USD [29], the LN is carving out a space for itself in the digital payments landscape.

\begin{figure}[ht] 
	\centering
        \includegraphics[width=0.5\textwidth]{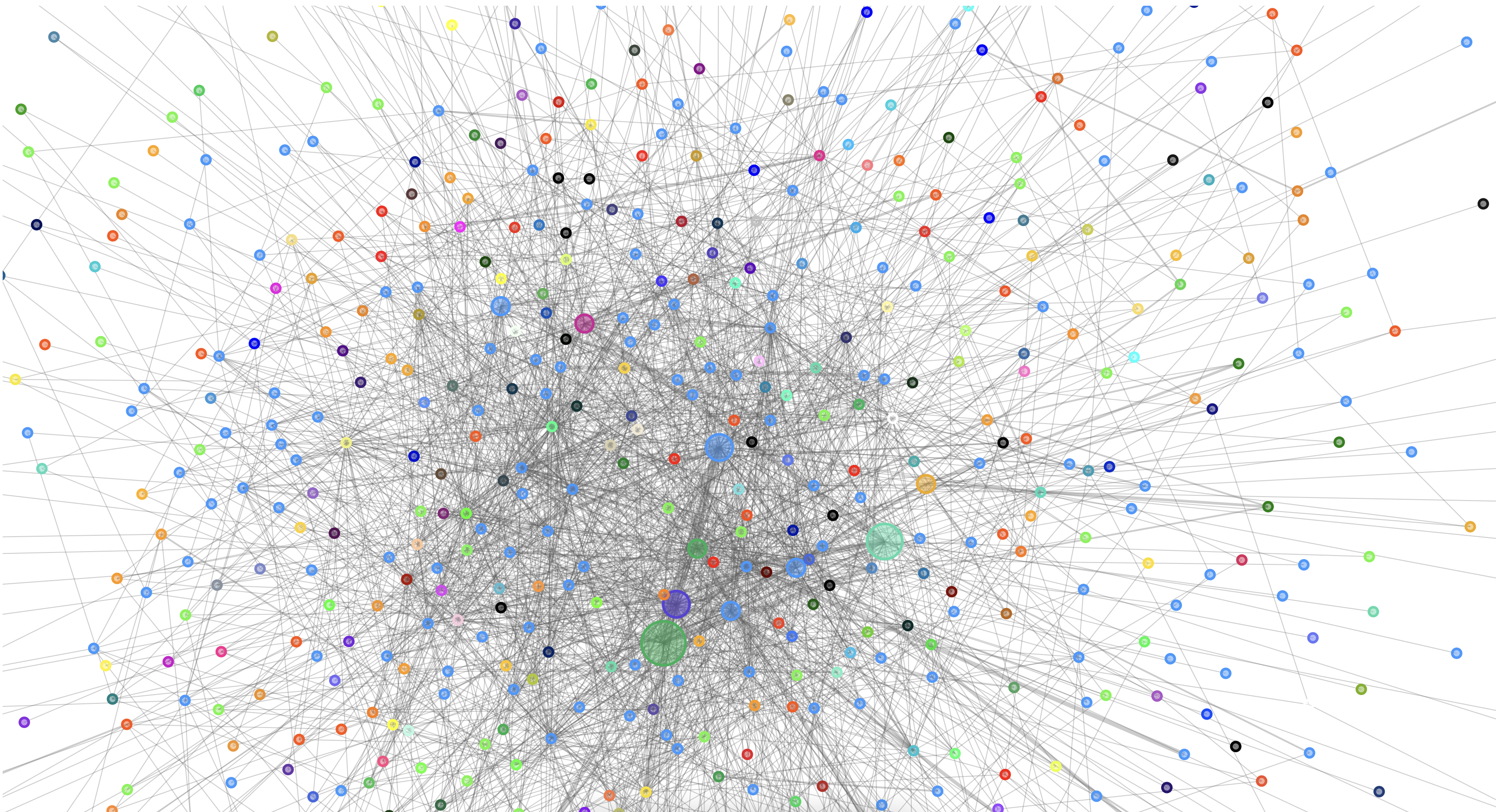}
	\captionsetup{justification=raggedright,singlelinecheck=false}\caption{Visualization of Lightning Network nodes. Source: 1ml, \href{https://1ml.com/visual/network}{https://1ml.com/visual/network.}}
	\label{fig:tcanther}
\end{figure}

\begin{figure}[ht] 
	\centering
        \includegraphics[width=0.5\textwidth]{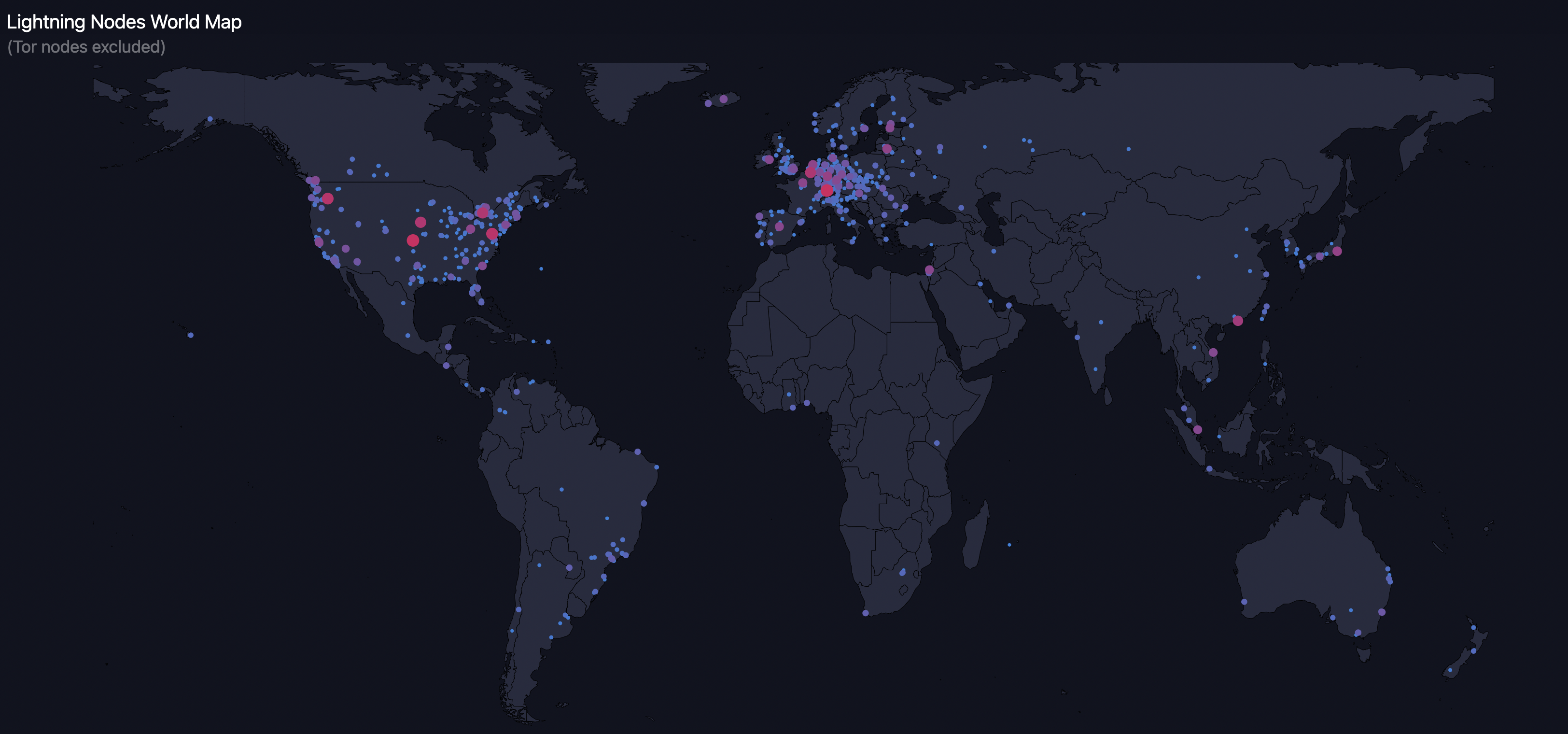}
	\captionsetup{justification=raggedright,singlelinecheck=false}\caption{Global map of Lightning Network nodes. Source: Mempool, \href{https://mempool.space/graphs/lightning/nodes-map}{https://mempool.space/graphs/lightning/nodes-map}.}
	\label{fig:tcanther}
\end{figure}

However, studies have revealed an inverse relationship between payment volume and success rate, suggesting that larger transactions face a lower probability of successful routing. This limitation is primarily due to liquidity issues, where only a fraction of the nodes can be reached for higher-value transfers [5, 6]. The network also suffers from information staleness regarding node availability and connectivity, particularly with IP and Tor addresses, which can affect routing success and network reliability [30].

From a structural perspective, the network exhibits signs of centralization, with a small number of nodes attracting a disproportionate amount of the transaction volume [31]. This skew in the network's layout has implications for its decentralization ethos. Metrics such as the Gini coefficient and the Nakamoto coefficient have shown an increase in inequality and a trend towards a more centralized core-periphery network structure over time [32, 43]. This centralization raises concerns about potential security threats, such as network fragmentation and split attacks, which could disrupt the integrity of the network.

Despite these challenges, adoption continues to grow, with more users and merchants integrating Lightning Network capabilities into their services. Innovations like Atomic Multipath Payments (AMP) are being explored to address the issue of routing larger payments by splitting them into smaller increments [33]. As the network matures and developers continue to refine its mechanisms, the likelihood of successfully routing payments is expected to improve.

Our paper acknowledges the current state of centralization within the LN and proposes a system that embraces a degree of centralization to enhance efficiency and reliability. While the Lightning Network's ethos is rooted in decentralization, our approach aims to provide a pragmatic solution for businesses seeking to leverage the network for their payment processing needs. By offering a centralized structure within the broader decentralized network, we aim to maintain objectivity and contribute to the network's evolution without undermining the core principles that underlie the Lightning Network's value proposition.

\subsection{2.2 Current State of the Payment Processing} 

We identify and discuss existing payment processing networks that are prevalent in the enterprise landscape. These systems form the backbone of current financial transactions in business, each with distinct characteristics in terms of transaction speeds and fee structures.

\textbf{Credit Cards}: These traditional credit card payment systems, including Visa and Mastercard, are renowned for their high transaction speeds, handling around 2,000 transactions per second with the capability to surge to several thousand during peak times [16, 17]. However, they are also known for their high fees, with Visa charging between 1.4\%-2.5\% plus a 0.14\% assessment fee, and Mastercard charging between 1.5\%-2.6\% plus a 0.1375\% assessment fee. American Express and Discover also fall into similar brackets, with their fees ranging from 1.55\%-3.5\%, accompanied by comparable assessment fees [34].

\textbf{PayPal}: A popular choice for online cross-border transactions, with almost 70\% of such buyers preferring PayPal [35]. It provides high transaction speeds but imposes significant fees, including 2.99\% plus a fixed fee for standard credit and debit card payments and 3.49\% plus a fixed fee for other commercial transactions [36].

\textbf{Western Union}: Known for its utility in domestic and international money transfers, Western Union typically completes domestic transfers within 24 hours and international transfers in one to five business days [37].  Outgoing wire transfer fees through banks typically range from \$20-\$35 for domestic transfers and \$35-\$50 for international transfers. Incoming wire transfers may incur fees ranging from \$0 to \$16 [38].

\textbf{Traditional Wire Transfers}: Often used for both domestic and international transfers, traditional wire transfers can take up to three days domestically, and potentially longer for international transfers [39]. The fees for these transfers vary, with outgoing domestic fees ranging from \$20-\$35, and international transfers costing between \$35-\$50. Incoming wire fees can also be charged, typically ranging up to \$16 [40].

\textbf{Automated Clearing House (ACH)}: ACH payments usually take 1-3 working days to be available in the recipient's account [41]. The fee structure for ACH processing varies, with some processors charging a flat rate per transaction (typically \$0.25 to \$0.75), while others charge a percentage fee, ranging from 0.5\% to 1\% per transaction [42].

It is evident that each traditional payment processing method has its limitations, particularly in terms of fees and transaction speeds. The associated costs often scale in terms of transaction amount and quantity. In contrast, we propose the Lightning Network as a competitive alternative. The Lightning Network stands out for its millisecond settlement speeds and minimal transaction fees, which are typically just a few cents [25]. This innovative approach offers a compelling solution for businesses looking to processes transactions of any size.


\section{3. Related Work}

\subsection{3.1 Spider}

In addressing the challenges of fast payments at scale within payment channel networks (PCNs), the work of Sivaraman et al. on the Spider routing solution presents a significant advancement. Spider innovatively "packetizes" transactions and employs a multi-path transport protocol to facilitate high-throughput routing in PCNs, effectively managing liquidity and balancing channel utilization. By breaking down transactions into smaller packets and using a multi-path approach, Spider can complete large transactions over time, even on channels with low capacity, and ensures fairness across payment flows. This method demonstrates a considerable improvement over traditional routing schemes, requiring substantially less on-chain rebalancing and achieving near-total throughput on balanced traffic demands [44].

While Spider addresses the critical issue of efficient routing in PCNs by optimizing the use of existing channels and reducing the need for on-chain rebalancing, our research explores a different dimension of scalability and efficiency in the Lightning Network. The quadratic growth in channel count, which is a fundamental limitation in fully meshed PCNs, remains unaddressed by Spider. As the number of clients in a network grows, the number of required channels increases quadratically, presenting a scalability bottleneck for large-scale adoption.

Our proposed dual-node closed payment network model seeks to overcome this limitation by introducing a topology that ensures linear scaling of channel requirements (O(n)) through a star graph configuration with two central hubs. This design not only simplifies channel management by reducing the total number of channels needed but also enhances privacy and security for enterprise-level applications. Moreover, our approach focuses on the closed network aspect, where all nodes connect exclusively within a controlled ecosystem, contrasting with the open and decentralized nature of existing PCNs where Spider operates.

\subsection{3.2 Raiden}

The Raiden Network, constructed atop the Ethereum blockchain, brings forth the versatility of ERC20 tokens through its support of smart contracts, enabling a wide array of decentralized applications. Conversely, our proposed payment network model is fundamentally anchored to the Bitcoin blockchain, a platform with a distinct security paradigm and a primary focus on facilitating peer-to-peer currency transactions. This focus on Bitcoin provides a stable and well-established foundation for our network model, which is particularly attractive to enterprises seeking a reliable digital currency with widespread recognition [45].

Smart contracts are integral to Raiden's operation, managing channel lifecycles and facilitating token transfers. Our model diverges from this approach, as the Bitcoin network does not inherently support complex smart contracts. Instead, we employ either manual channel management or an automated solution configured to the Bitcoin protocol. This simplification aligns with the needs of businesses that may prefer direct control over channel management or seek an automated system that operates within the constraints of Bitcoin's scripting capabilities.

The topology of our network is designed to streamline payment routing, utilizing a star graph with two central nodes to ensure guaranteed and efficient routes for transactions. This structure starkly contrasts with Raiden's multi-hop network, which requires intricate routing algorithms to navigate a web of channels. By simplifying the payment routes, our model provides a more predictable and reliable transaction flow, essential for enterprise applications where the certainty of payment delivery is critical.

Liquidity management is another area where our model offers significant advantages. While Raiden channels are restricted by the tokens locked in smart contracts, potentially complicating liquidity distribution, our dual-node configuration includes robust rebalancing capabilities. Central nodes facilitate liquidity management, ensuring channels are adequately funded to handle transactions in both directions. This proactive rebalancing minimizes delays and avoids the need for frequent, costly on-chain transactions, making it ideal for handling the larger transaction volumes typically seen in B2B environments.

Our network model is expressly designed for enterprise use, focusing on B2B transactions where stability, security, and consistent throughput are non-negotiable. It caters to businesses that require a high degree of predictability in their payment processing infrastructure, a stark contrast to Raiden's micro-transaction orientation that targets consumer-level interactions and decentralized applications.

Lastly, the fee structure in our model is crafted to offer predictability and control. Unlike Raiden's competitive fee market for intermediaries, which can lead to variable transaction costs, our closed network design allows for a more consistent fee schedule. This predictability in costs is crucial for businesses that need to budget and forecast expenses accurately, ensuring that financial planning is not disrupted by fluctuating fee structures.

\subsection{3.3 Sidechains}

Bitcoin's scalability challenges have led to the proposal and development of various sidechain solutions, each with its unique approach to enhancing transaction speed, efficiency, and privacy. Three notable sidechain solutions are the Liquid Network, Drivechain, and Statechains [46, 47, 48].

\textbf{Liquid Network: }
The Liquid Network is a federated sidechain developed by Blockstream, which allows for faster Bitcoin transactions with enhanced privacy features through Confidential Transactions. It is primarily targeted at exchanges and financial institutions that require rapid settlement times. The Liquid Network operates with a fixed set of validators that are responsible for the creation of blocks, making it more centralized than Bitcoin's main chain [46].

\textbf{Drivechain: }
Drivechain is a sidechain proposal that allows Bitcoins to be moved to a separate blockchain with different rules and then moved back to the main Bitcoin blockchain. This enables the creation of various sidechains, each tailored to specific use cases or scalability solutions, without affecting the main chain. Drivechain relies on miners to vote on the movement of funds between the main chain and sidechains, introducing a new dimension of miner involvement in network operations [47].

\textbf{Statechains: }
Statechains are a layer-2 scaling solution that allows users to transfer the state of a Bitcoin UTXO off-chain. This is achieved without requiring trust in an intermediary, by transferring private key information in a secure manner. Statechains complement the Lightning Network by providing a different approach to off-chain transactions, focusing on UTXO ownership rather than payment channels [48].
\\ \\
While these sidechain solutions offer potential benefits, there are key differences when compared to our proposed dual-node closed payment network on the Lightning Network:

\begin{itemize}
    \item \textbf{Complexity}: Operating a sidechain requires managing the pegging process between the main chain and the sidechain, which adds complexity to the transaction process. Our approach avoids this by working within the existing Lightning Network infrastructure, simplifying operations for businesses.
    \item \textbf{Tokenomics}: Sidechains can introduce new tokenomics, as assets on the sidechain may differ from those on the main chain. Our approach leverages the native Bitcoin currency within the Lightning Network, avoiding the introduction of new tokens or assets.
    \item \textbf{Security}: The security model of sidechains can differ from that of the main Bitcoin blockchain, potentially introducing new security considerations. By staying on the Lightning Network, our approach benefits from the security of the underlying Bitcoin protocol.
\end{itemize}

Given these considerations, we opt for the operational simplicity and security of staying on the Lightning Network rather than running a separate sidechain. This allows businesses to leverage the proven infrastructure of the Lightning Network without the need for additional mechanisms such as pegs, while still achieving the desired scalability and efficiency for client-to-client transactions.


\section{4. Terminology}

For the sake of clarity in this discussion, we define the following terms:

\subsection{4.1 Closed Payment Network}
A closed payment network is a type of payment system characterized by restricted participation and private transaction channels. It is designed to serve a specific group of users, often within a particular organization or consortium. The key attributes of a closed payment network include:

\begin{itemize}
  \item Restricted Access: Only authorized participants are allowed to join the network, ensuring a controlled environment for transaction processing.
  \item Privacy: Transactions within the network are not publicly broadcast, providing an additional layer of privacy for participants. In terms of the Lightning Network, the only public transactions are those committed to the mainnet chain, such as opening and closing channels.
  \item Customization: The network can be tailored to meet the specific needs of its facilitator or users, including compliance with regulatory requirements and integration with other systems.
\end{itemize}

\subsection{4.2 Private Payment Facilitator}
A private payment facilitator refers to an entity that operates a payment network on behalf of a closed group of clients. The facilitator is responsible for the management and maintenance of the network's infrastructure. Characteristics of a private payment facilitator include:

\begin{itemize}
  \item Service Provider: Acts as an intermediary that processes payments, manages liquidity, and ensures the smooth operation of the payment network.
  \item Operational Control: Maintains control over the network's rules, policies, and technical specifications to optimize for desired outcomes.
  \item Cost Structure: Has the flexibility to determine the network's fee model, ranging from traditional transaction fees to a fee-less structure where operational costs are subsidized as part of a broader service agreement.
\end{itemize}

\subsection{4.3 Client}
A client refers to an individual or entity that engages the services of a private payment facilitator to process payments within the closed payment network. Clients are the end-users for whom the network's transactional capabilities are provided. In the network topology, clients are represented by client nodes, which are the terminal points of the network through which transactions are initiated and received.


\section{5. Closed Network Topology}

In the design of a closed payment network leveraging the Lightning Network, the network topology plays a crucial role in determining both the initial setup cost and the ongoing maintenance expenses, as well as the operational complexity. We consider two distinct network topologies: the interconnected mesh topology and the centralized hub-and-spoke topology.

\subsection{5.1 Complete Graphs and Mesh Topology}

\begin{figure}[ht] 
	\centering
        \includegraphics[width=0.2\textwidth]{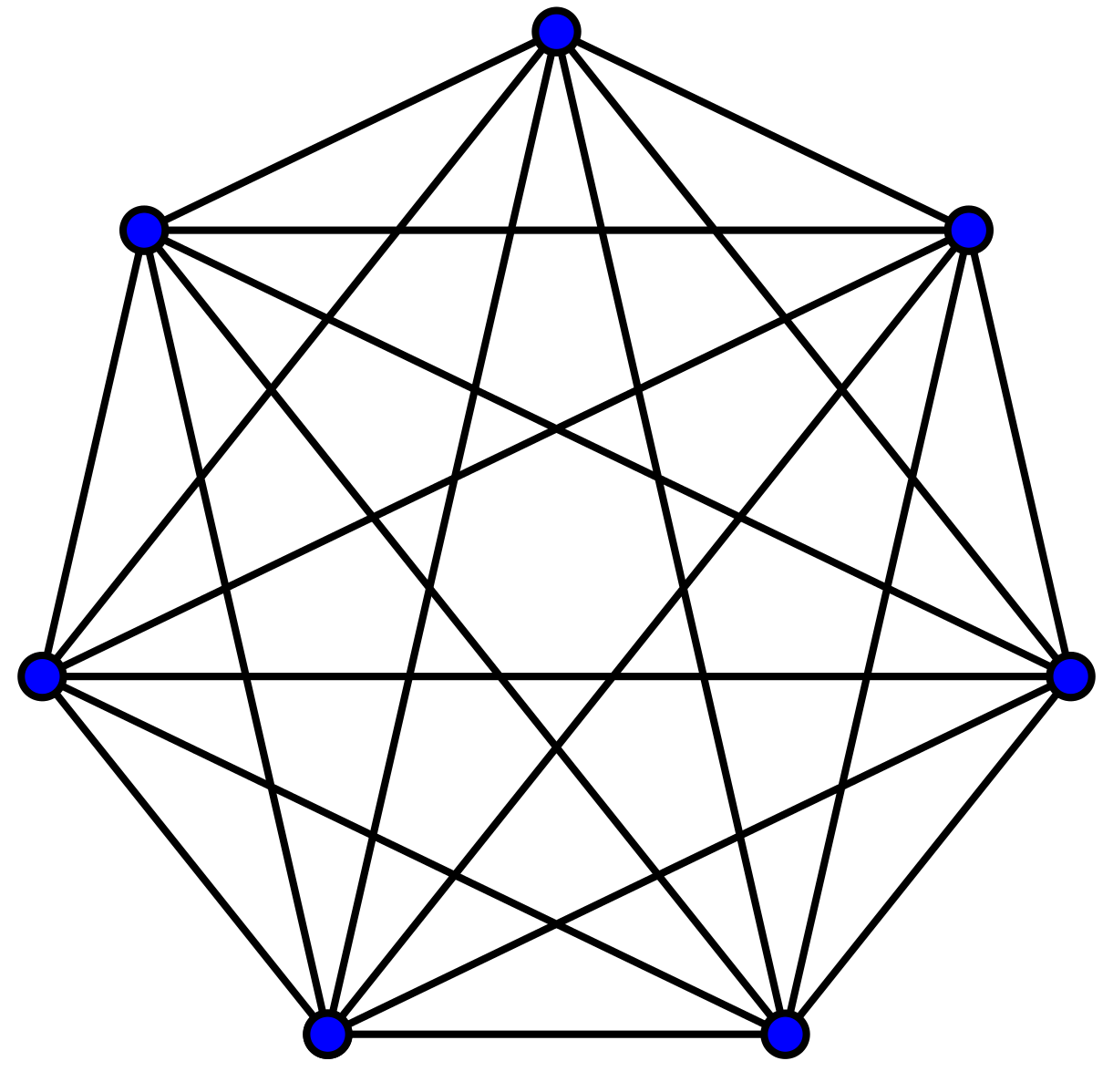}
	\captionsetup{justification=raggedright,singlelinecheck=false}\caption{Complete graph / interconected mesh. Source: Wikipedia, \href{https://commons.wikimedia.org/wiki/File:Complete_graph_K7.svg}{https://commons.wikimedia.org/wiki/File:Complete\\\_graph\_K7.svg}.}
	\label{fig:tcanther}
\end{figure}

The interconnected mesh topology is represented by a complete graph, denoted as \( K_n \), where \( n \) signifies the number of nodes, each corresponding to a client within the network. A complete graph is defined by its edge density, with the number of edges \( E \) given by the formula:

\begin{equation}
	E = \frac{n(n - 1)}{2}
\end{equation}

This equation reveals that the number of edges grows quadratically with the number of nodes, representing a fully connected network where each node is directly linked to every other node. The high edge density of a complete graph ensures that there are multiple direct paths for routing payments, which enhances the network's resilience by providing abundant alternative routes in case some channels become unavailable.

Despite the robustness conferred by this topology, it becomes increasingly impractical as the network scales. The cost of opening a new channel involves not only the transaction fees on the blockchain but also the opportunity cost of the capital that becomes locked in each channel. As the network expands, the requirement for each node to establish direct channels with all other nodes results in a quadratic increase in the total number of channels. This leads to a surge in the aggregate costs of transaction fees for channel creation and closure, as well as a substantial amount of capital being tied up across the network.

Moreover, the complexity of managing such a dense network becomes a significant operational challenge. Each node must monitor and manage \( n - 1 \) channels, which entails not only keeping track of the state of each channel but also ensuring that there is sufficient liquidity for transactions to proceed smoothly. The administrative burden of managing these channels includes handling channel lifecycle events such as opening, closing, and rebalancing, which requires constant vigilance and active management.

The rebalancing of channels in a complete graph is particularly complex, as it involves considering the liquidity state of every possible channel path. If a node needs to rebalance its channels, it must potentially engage in multiple transactions across various paths to redistribute liquidity. This introduces the issue of an algorithmic search to determine the optimal path. Each transaction executed during the rebalancing process alters the liquidity profile of the channels it traverses, thereby influencing the subsequent route calculations for the following transactions; this cascading effect compounds the complexity, as each rebalance operation must account for the dynamic liquidity landscape shaped by the preceding transactions.

The computational complexity for determining the optimal rebalancing transactions can become prohibitive for  private payment facilitators as the network grows.

\subsection{5.2 Star Graphs and Hub-and-Spoke Topology}

\begin{figure}[ht] 
	\centering
        \includegraphics[width=0.4\textwidth]{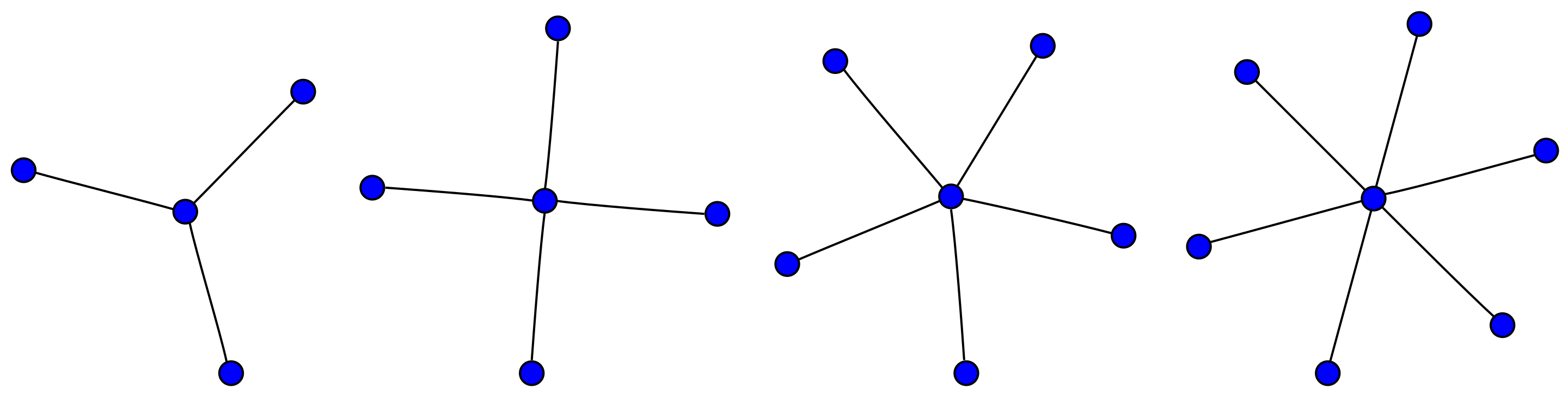}
	\captionsetup{justification=raggedright,singlelinecheck=false}\caption{Star graph / hub-and-spoke. Source: Wikipedia, \href{https://commons.wikimedia.org/wiki/File:Star_graphs.svg}{https://commons.wikimedia.org/wiki/File:Star\_graphs.svg}.}
	\label{fig:tcanther}
\end{figure}

Conversely, the centralized hub-and-spoke topology is represented by a star graph, where a central hub node is connected to all other nodes. The number of edges \( E \) is linearly proportional to the number of nodes \( n \), excluding the central hub:

\begin{equation}
	E = n - 1
\end{equation}

In this low-density network, the addition of a node only necessitates the creation of a single new channel, which is more resource-efficient. The star graph topology minimizes the number of on-chain transactions for channel management, reducing both the associated fees and the waiting time for confirmations on the mainnet. Each additional node in the network requires only a single on-chain transaction to connect to the central hub, as opposed to multiple transactions in a mesh network.

Moreover, the maintenance of channels in a star graph involves monitoring and managing fewer channels overall, which simplifies the operational requirements. The central hub can efficiently allocate resources to monitor the health and liquidity of each channel, making it easier to identify and address potential issues such as channel closures or rebalancing needs.

In a centralized hub-and-spoke model, liquidity is concentrated at the central hub, which can be advantageous for a closed payment network. The central node can manage liquidity more effectively, ensuring that sufficient funds are available to facilitate transactions in any direction. This concentration simplifies the process of liquidity provisioning and can lead to more predictable transaction costs and improved capital efficiency.

Rebalancing in a star graph topology involves fewer potential paths, reducing the complexity of finding efficient rebalancing routes. The central node can act as a counterparty for all rebalancing operations, which streamlines the process and reduces the computational overhead involved in identifying rebalancing opportunities.

The star graph topology inherently offers better scalability for a growing closed payment network. As the network expands, the addition of new nodes does not disproportionately increase the complexity nor cost of  maintenance. This makes the hub-and-spoke model particularly attractive for private payment facilitators that anticipate a growing number of internal transactions or expanding client bases.

\subsection{5.3 Tradeoffs}

The adoption of a star graph topology in the closed payment network introduces certain tradeoffs when compared to a mesh network. A primary consideration is that transactions between client nodes necessitate two Lightning Network transactions: one to the central routing node and another from the central routing node to the destination node. This two-step process is a departure from the direct node-to-node transactions in a mesh topology.

Despite this, the tradeoff is generally favorable when considering the operational efficiencies gained. The star graph topology significantly reduces the total number of channels required, which in turn lowers the costs and complexities associated with opening and closing channels. In a mesh network, the quadratic growth in the number of channels with each additional node leads to increased fees and capital locked in liquidity, which can become prohibitive as the network scales.

Transaction times within the Lightning Network are typically measured in milliseconds [25], which means the additional time incurred by the two-step process is minimal. Over time, the cumulative time cost is likely to be less significant compared to the time and fees associated with managing a larger number of channels in a mesh topology, which requires lengthy mainnet transactions.

Furthermore, the fee structure within the closed payment network can be optimized by the private payment facilitator. Since the facilitator controls the central routing node, they have the discretion to set channel fees, potentially reducing or eliminating them. This contrasts with a mesh network, where transactions may incur fees from multiple nodes if routed through two or more nodes to reach the destination, especially if liquidity constraints necessitate longer routing paths.


\section{6. Central Routing Node}

In the context of a closed payment network, a central routing node serves as the core for payment processing and channel management.

\subsection{6.1 Definition and Role}

A central routing node, within the framework of a closed payment network, is a specialized node that acts as the primary intermediary for processing transactions between clients. This approach is analogous to traditional centralized financial networks, where a central authority or clearinghouse facilitates transactions between various parties. 

The central routing node is responsible for:

\begin{itemize}
    \item Establishing and maintaining payment channels with all nodes in the network, thereby reducing the need for each node to manage multiple channels.
    \item Facilitating the routing of transactions across the network, ensuring that payments are executed promptly and efficiently.
    \item Managing the liquidity of channels to guarantee that sufficient funds are available for transaction settlement in any direction, thus minimizing delays and the need for frequent rebalancing.
    \item Monitoring the health and stability of the network, proactively addressing potential issues such as channel congestion or imbalances before they impact transaction flows.
\end{itemize}

By centralizing transaction processing, the closed payment network can achieve a level of efficiency and predictability that makes the Lightning Network a feasible alternative to traditional payment processors.

\begin{figure}[ht] 
	\centering
        \includegraphics[width=0.4\textwidth]{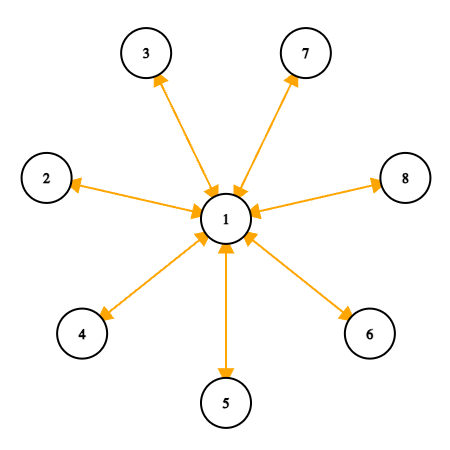}
	\captionsetup{justification=raggedright,singlelinecheck=false}\caption{Central routing node. Source: Self-generated by Jeffy Yu}
	\label{fig:tcanther}
\end{figure}

\subsection{6.2 Channel and Liquidity Management}

In a centralized routing node model, the central node assumes the role of managing the network's liquidity and channels, unlike a decentralized mesh topology where each node must manage and monitor its own set of channels.

\textbf{Theorem 1: Channel Management Efficiency.} Let \( C_i \) be the set of channels for a node \( i \) in a mesh network with \( n \) nodes. The total number of channel queries \( Q_m \) in a mesh network is given by the sum of the number of channels for each node, which can be expressed as:

\[
Q_m = \sum_{i=1}^{n} |C_i|
\]

In a centralized model with a single routing node \( R \), the number of channel queries \( Q_c \) reduces to the number of channels connected to \( R \), denoted as \( C_R \):

\[
Q_c = |C_R|
\]

Since \( C_R = n - 1 \) for a star topology, where \( n \) includes the central node, we have:

\[
Q_c = n - 1
\]

\textbf{Corollary:} In a mesh network, \( Q_m \) grows quadratically with \( n \), while in a centralized model, \( Q_c \) grows linearly, proving the centralized model's superior efficiency in channel management.

\textbf{Theorem 2: Liquidity Management Optimization.} Assume a network with a central routing node \( R \), and payment nodes \( P_1, P_2, ..., P_{n-1} \). Each channel \( C_{iR} \) between \( P_i \) and \( R \) has two sides, with \( l_{iR} \) liquidity on the side of \( P_i \) and \( l_{Ri} \) liquidity on the side of \( R \). The objective of liquidity management is to minimize the total rebalancing cost \( T \) across all channels \( C_{iR} \), which can be formulated as:

\[
\min T = \sum_{i=1}^{n-1} \text{cost}(\text{rebalance}_{iR})
\]

where \( \text{cost}(\text{rebalance}_{iR}) \) represents the cost of rebalancing channel \( C_{iR} \), incorporating both on-chain fees and off-chain routing fees.

\textbf{Corollary:} The centralized node can approach this optimization problem by leveraging its global view of the network's liquidity state. This allows for a comprehensive solution that minimizes the total rebalancing cost, which is more efficient than addressing individual rebalancing issues in isolation. The central node can execute a series of rebalancing transactions that collectively optimize network liquidity, as opposed to decentralized nodes that would need to negotiate and execute rebalancing transactions on a per-channel basis without full visibility of the network's state.

To contrast this with the decentralized mesh topology, consider the rebalancing problem for an individual node \( P_i \) with channels \( C_{ij} \) to other nodes \( P_j \). In this case, the rebalancing cost minimization for node \( P_i \) is:

\[
\min T_i = \sum_{j=1, j \neq i}^{n-1} \text{cost}(\text{rebalance}_{ij})
\]

In a decentralized setting, each node \( P_i \) must independently solve this problem, lacking the global perspective on liquidity distribution. This can lead to suboptimal rebalancing transactions that may temporarily alleviate liquidity constraints for \( P_i \) but could exacerbate imbalances elsewhere in the network.

Furthermore, the decentralized rebalancing process requires synchronization and communication between multiple nodes, which can introduce additional complexity and inefficiency. In contrast, the central routing node can rebalance channels \( C_{iR} \) by pushing liquidity from one side to the other as needed, using its comprehensive knowledge of the network's liquidity state to execute the most cost-effective transactions.

\subsection{6.3 Linear Programming Optimization}

A linear programming approach can be used to determine the optimal amounts of liquidity to move between the central node and each of the payment nodes in order to maintain the desired liquidity levels while minimizing costs. To implement this optimization using linear programming, we define: \\

- \( x_i \): The amount of liquidity to move from \( R \) to \( P_i \) for rebalancing.
- \( c_i \): The cost per unit of liquidity moved in channel \( C_{iR} \).
- \( L_{\text{min}} \): The minimum required liquidity in any channel.
- \( L_{\text{available}} \): The total liquidity available at node \( R \) for rebalancing. \\

The linear programming formulation is as follows. \\

Objective function:
\[
\min Z = \sum_{i=1}^{n-1} c_i \cdot y_i
\]

Subject to:
\[
l_{iR} + x_i \geq L_{\text{min}}, \quad \forall i \in \{1, 2, ..., n-1\}
\]
\[
l_{Ri} - x_i \geq L_{\text{min}}, \quad \forall i \in \{1, 2, ..., n-1\}
\]
\[
\sum_{i=1}^{n-1} x_i \leq L_{\text{available}}
\]
\[
-x_i \leq l_{iR}, \quad \forall i \in \{1, 2, ..., n-1\}
\]
\[
x_i \leq l_{Ri}, \quad \forall i \in \{1, 2, ..., n-1\}
\]
\[
y_i \geq x_i, \quad \forall i \in \{1, 2, ..., n-1\}
\]
\[
y_i \geq -x_i, \quad \forall i \in \{1, 2, ..., n-1\}
\]

where \( y_i \) are auxiliary variables introduced to linearize the objective function. Solving this linear programming problem yields the optimal rebalancing amounts \( x_i \) for each channel, minimizing the total rebalancing cost \( T \) while ensuring liquidity requirements are met.

\subsection{6.4 Rebalancing Strategies}
Effective rebalancing strategies are essential for the central routing node to ensure liquidity and maintain network efficiency. Automated systems can be implemented to regularly query the central routing node's API, monitoring the state of each channel's liquidity. When the liquidity of any given channel falls below a predefined threshold, the system can trigger a rebalancing operation.

One method to execute rebalancing is through linear programming, as detailed in the previous subsection. This approach provides an optimal solution for redistributing liquidity across channels by minimizing the total rebalancing cost while adhering to liquidity constraints.

Circular rebalancing also presents a viable strategy, allowing funds to be shifted within the network in a loop to balance the channels without the need for external transactions. However, this requires specific alterations to the network topology, which will be discussed in a subsequent section on a dual-node approach. 

\subsection{6.5 Internal Fee Structure}
Within the closed payment network operated by a private payment facilitator, the fee structure plays a critical role in the network's economic viability and the value proposition offered to clients. The Lightning Network Daemon (LND) provides two primary fee parameters for channel management: the base fee and the fee rate [65]. The base fee, denoted as \texttt{base\_fee\_msat}, is a fixed fee charged for each forwarded HTLC (Hashed Time-Locked Contract), expressed in milli-satoshi. In contrast, the fee rate, indicated by \texttt{fee\_rate\_ppm} (parts per million), is a variable fee that scales with the value of the HTLC.

\begin{verbatim}
Setting base fee and fee rate:
lncli updatechanpolicy --base_fee_msat 1000 
--fee_rate_ppm 300 <channel_id> [65]
\end{verbatim}

In a closed payment network, the facilitator has the discretion to set these fees to optimize the balance between covering operational costs and providing competitive transaction fees to clients. Given that the primary goal is to process transactions efficiently and affordably, the facilitator may choose to set a minimal base fee, or even zero, as a means to incentivize network usage.

The private payment facilitator can potentially absorb the costs of channel management and rebalancing as part of a broader business model. This approach can be particularly advantageous in fostering a high volume of internal transactions, where the marginal cost of an additional payment is low, and the overall increase in network activity contributes to the bottom line through other value-added services.


\section{7. Dual Central Node Approach}

\subsection{7.1 Architectural Overview}

In the dual-node configuration of the closed payment network, a central routing node is complemented by a secondary, dormant node. The central routing node is tasked with the primary functions of transaction routing and channel management. It maintains direct channels with all client nodes within the network, facilitating the execution of transactions.

The dormant node remains inactive under normal network conditions, holding a mirrored set of channels to those of the central routing node, albeit with minimal liquidity. It also has a direct channel to the central routing node. 

Its primary functions are to participate in the network's rebalancing activities and to provide redundancy for fault tolerance. Circular rebalancing is enabled by the dormant node as it completes the circuit between the client nodes and the central routing node. This connectivity allows the network to reallocate liquidity imbalances by routing funds through a loop that begins and ends at the central routing node.

Circular rebalancing is inherently more efficient in a closed network as it allows for liquidity adjustments without the need for external funds or on-chain transactions. The liquidity remains within the system, thereby conserving the total network value and minimizing dependency on external liquidity sources, which can incur additional costs and delays.

Additionally, transaction costs are minimized by reducing the number of necessary on-chain transactions. Since the rebalancing occurs entirely off-chain within the closed loop, the network avoids the fees and waiting times associated with mainnet blockchain confirmations.

The connection topology between the central routing node, the dormant node, and the client nodes is constructed to form a star graph with two central hubs. Each client node is connected to both the central routing node and the dormant node through manually established channels. 

\begin{figure}[ht] 
	\centering
        \includegraphics[width=0.4\textwidth]{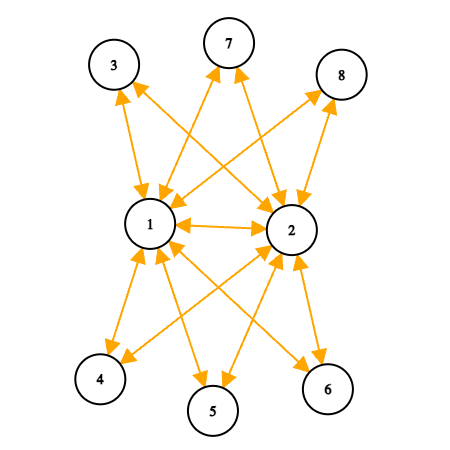}
	\captionsetup{justification=raggedright,singlelinecheck=false}\caption{Dual central node approach. Source: Self-generated by Jeffy Yu}
	\label{fig:tcanther}
\end{figure}

\subsection{7.2 Modeling of Circular Rebalancing}

Circular rebalancing within the dual-node system is a process that allows for the redistribution of liquidity in a closed loop, ensuring that each channel within the network maintains optimal liquidity levels for transaction processing. The mathematical model for circular rebalancing is based on the conservation of value principle, where the total amount of liquidity within the network remains constant, and only its allocation across different channels changes.

Let us denote the liquidity in a channel from node \(i\) to node \(j\) as \(L_{ij}\). For circular rebalancing, we consider a sequence of channels forming a closed loop that includes the central routing node (CR), the dormant node (DN), and a subset of client nodes (\(P_1, P_2, \ldots, P_k\)) within the network. The objective is to perform a series of transactions that redistribute liquidity without the need for external funds.

The conservation of liquidity in a circular rebalancing operation can be represented by the following system of equations:

\begin{equation}
    L_{CR,P_1}^{\prime} = L_{CR,P_1} - x_1
\end{equation}
\begin{equation}
    L_{P_1,P_2}^{\prime} = L_{P_1,P_2} + x_1 - x_2
\end{equation}
\begin{equation}
    \vdots
\end{equation}
\begin{equation}
    L_{P_{k-1},P_k}^{\prime} = L_{P_{k-1},P_k} + x_{k-1} - x_k
\end{equation}
\begin{equation}
    L_{P_k,DN}^{\prime} = L_{P_k,DN} + x_k - y
\end{equation}
\begin{equation}
    L_{DN,CR}^{\prime} = L_{DN,CR} + y
\end{equation}
where:
- \(L_{ij}^{\prime}\) is the liquidity in the channel after rebalancing,
- \(x_i\) is the amount of liquidity transferred from node \(i\) to node \(i+1\),
- \(y\) is the amount of liquidity transferred from the dormant node back to the central routing node, completing the circle.

The objective of circular rebalancing is to find a set of transfer amounts \(\{x_1, x_2, \ldots, x_k, y\}\) that satisfy the liquidity requirements of each channel while minimizing the number of transactions and the amount of liquidity moved. This can be formulated as an optimization problem:

\begin{equation}
    \min \sum_{i=1}^{k} |x_i| + |y|
\end{equation}

Subject to:
\begin{equation}
    L_{ij}^{\prime} \geq L_{\text{min}}, \quad \forall i, j
\end{equation}
\begin{equation}
    \sum_{i=1}^{k} x_i = y
\end{equation} \\

where \(L_{\text{min}}\) is the minimum required liquidity in any channel to ensure efficient transaction processing.

The optimization problem can be solved using linear programming techniques or other numerical methods suitable for the network's scale and complexity. The resulting values of \(\{x_1, x_2, \ldots, x_k, y\}\) dictate the exact transactions to be executed in the rebalancing process, ensuring that each channel is sufficiently funded to handle the expected transaction volume.

\subsection{7.3 Closed Network Integrity}

In a closed payment network, it is critical to maintain the network's integrity and privacy. To this end, the dual-node system eschews the autopilot channel management found in open Lightning Network implementations in favor of the manual setting. This management ensures that the network remains isolated from external transactions, preserving liquidity for the exclusive use of network participants and upholding the network's privacy.

The management is characterized by the following key operations:

\begin{itemize}
    \item \textbf{Channel Establishment:} Channels between new client nodes and the central nodes are opened manually by the private payment facilitator.
    \item \textbf{Channel Monitoring:} Continuous monitoring of channel states through the central routing node API allows for the proactive management of liquidity levels and channel capacity.
    \item \textbf{Channel Adjustment:} Manual interventions are performed to rebalance or adjust channel liquidity.
    \item \textbf{Channel Closure:} When necessary, channels are manually closed by the private payment facilitator, releasing funds.
\end{itemize}

While the network does not automatically accept external connections, these operations will likely be automated by an algorithm or program that connects to the central routing node's API.

Moreover, the closed payment network ensures transaction privacy by keeping off-chain transfers invisible to the public. Only the channel openings and closings are recorded on the blockchain, while the actual transactions remain private within the network. The private payment facilitator may maintain an internal ledger for record-keeping, but this information is not exposed to the public ledger.

Since the network's structure is known and all nodes are under the facilitator's control, complex privacy protocols like onion routing are unnecessary. There's only one direct route for transactions between nodes, which simplifies the process and maintains privacy without additional measures.

\section{8. Fault Tolerance}

\subsection{8.1 Dormant Node}

Fault tolerance within the closed payment network is a critical aspect that ensures continuous operation and resilience against node failures. Having a only one central routing node creates a single point of failure for the entire network. The dormant node provides fault tolerance. It is configured to mirror the channel structure of the central routing node but remains inactive during normal network operations. The dormant node serves as a standby system that can rapidly assume the responsibilities of the central routing node in the event of a failure.

The mechanism for the dormant node to take over involves several steps. The private payment facilitator should continuously monitor the status and uptime of the primary central routing node, and activate this contingency during abnormalities. Upon detection of a failure in the central routing node—be it due to hardware malfunctions, software issues, or external disruptions—the network's protocols initiate a switchover process. This process entails the dormant node transitioning from its standby state to an active state, where it begins to process transactions and manage liquidity as the central routing node did.

For the switchover to be successful, the dormant node must have access to a significant amount of liquidity. This liquidity ensures that the dormant node can handle the transaction volume and maintain the necessary channel balances to facilitate uninterrupted payment processing. The private payment facilitator is responsible for pre-allocating this liquidity to the dormant node, ensuring that it is ready to be deployed instantly when needed. 

Optimally, the standby liquidity is stored in both the node wallet and the channels with the client nodes. If the funds are stored within the node wallet, the time cost of a mainnet transaction exists to replenish the channels with the client nodes before taking over operations. If it is stored as Bitcoin on chain, the time cost is doubled, since a mainnet transaction is needed to fund the node and then replenish channels for use.

The liquidity allocation should be strategically planned to mirror the active node's liquidity profile, allowing for a one-to-one correspondence in channel capacity and ensuring no degradation in service quality during the switchover.

\subsection{8.2 Multi-Tiered Redundancy}

To further enhance the fault tolerance of the closed payment network, additional dormant nodes can be incorporated, creating a multi-tiered redundancy architecture. This approach introduces layers of fallback options, where each additional dormant node serves as a contingency for both the central routing node and the preceding dormant nodes in the redundancy hierarchy.

The inclusion of a third central routing node, for instance, establishes a system with two levels of fallback. The primary central routing node conducts the network's transaction routing and channel management under normal conditions. If it encounters a failure, the first dormant node is activated to take over its duties. Should the first dormant node also fail or become compromised, the second dormant node is then activated, providing an additional layer of redundancy.

\begin{figure}[ht] 
	\centering
        \includegraphics[width=0.4\textwidth]{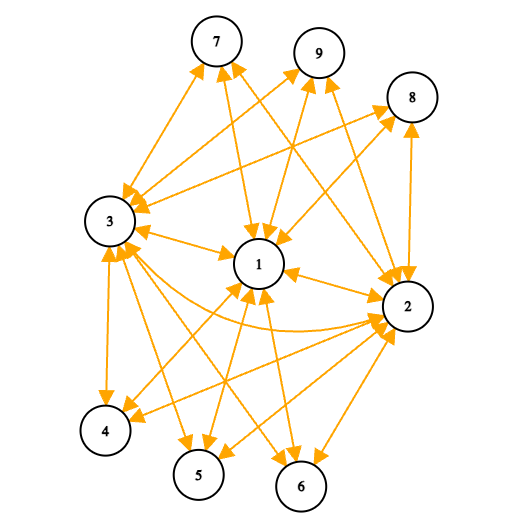}
	\captionsetup{justification=raggedright,singlelinecheck=false}\caption{Multi-tiered redundancy. Source: Self-generated by Jeffy Yu}
	\label{fig:tcanther}
\end{figure}

Each dormant node in the multi-tiered architecture must be provisioned with sufficient liquidity to fulfill the central routing node's functions upon activation. The liquidity distribution across the dormant nodes should be strategically managed to ensure that each node can maintain the network's operational integrity if called upon. 

This may involve periodic rebalancing between the dormant nodes to align their liquidity profiles with the evolving transaction patterns of the network. The private payment facilitator can bundle this in the circular rebalancing of the client node channels to have a single batch of transactions.

The activation of dormant nodes in a multi-tiered redundancy system can be managed through a combination of automated monitoring and manual oversight. Automated systems continuously assess the health and performance of the central routing node and the active dormant nodes, triggering alerts and initiating switchover protocols when predefined conditions are met. Manual oversight by the private payment facilitator's operations team provides an additional layer of control, allowing for human intervention in complex or ambiguous failure scenarios.


\section{9. Throughput Modeling}

\subsection{9.1 Methodology}
To quantitatively assess the throughput of our proposed dual-node Lightning Network payment system, a simple discrete-event simulation is created with the SimPy framework. The simulation's goal was to measure the network's transaction processing capacity over a one-month period. We set parameters such as transaction processing time, frequency and duration of rebalancing events, and the total number of nodes participating in the network. Each payment is broken down into two transactions, first to the central routing node, and then to the destination client node.

\subsection{9.2 Simulation Parameters}
\begin{itemize}
    \item \textbf{Transaction Processing Time}: 0.3 seconds, to simulate rapid transaction settlement on the Lightning Network. 
    \item \textbf{Rebalancing Interval and Duration}: Scheduled daily with a one-hour downtime, representing an overestimation for the simulation's purposes. In a production environment, the network is expected to either switch to a dormant node or complete rebalancing more swiftly.
    \item \textbf{Simulation Duration}: 2592000 seconds (one month), 15778800 (six months), and 31536000 (one year), providing a comprehensive view of the network's long-term throughput.
    \item \textbf{Node Configuration}: One central node for transaction processing and 100 payment nodes. Dormant node not modeled since time factor of rebalancing is only considered.
\end{itemize}

\subsection{9.3 Results}

\begin{table}[ht] 
	\caption{SimPy Simulation Results.}
	\centering
	\begin{tabular}{l l r}
		\toprule
		\multicolumn{3}{c}{Completed Payments} \\
		\cmidrule(r){1-3}
		Time & Nodes & Payments \\
		\midrule
		1 month & 100 & 2484000 \\
		6 months & 100 & 14904000 \\
		1 year & 100 & 29808000 \\
		\bottomrule
	\end{tabular}
	\label{tab:distcounts}
\end{table}

\subsection{9.4 Discussion}

The SimPy simulation results, as summarized in Table \ref{tab:distcounts}, reveal key insights into the throughput capacity of our dual-node Lightning Network payment system under various operational durations. The network successfully processed 2,484,000 payments in a month, 14,904,000 payments in six months, and 29,808,000 payments over a year with 100 nodes. These results demonstrate substantial scalability and robustness in handling a high volume of transactions over extended periods.

However, it is crucial to contextualize these results within the limitations and assumptions of the simulation:

\begin{enumerate}
    \item \textbf{Rebalancing Overestimation}: The daily rebalancing with one-hour downtime represents a conservative estimate. In a real-world scenario, efficient rebalancing strategies, such as switching to a dormant node or quicker rebalancing processes, could significantly reduce downtime, thereby increasing the total throughput.
    \item \textbf{Uniform Transaction Processing Time}: The simulation assumes a uniform processing time of 0.3 seconds per transaction. Actual transaction times may vary based on network load, node capabilities, and transaction complexity.
    \item \textbf{Maintenance and Operational Challenges}: Regular network maintenance, software updates, and potential technical issues were not simulated.
    \item \textbf{Scalability with More Nodes}: While the simulation was conducted with 100 nodes, adding more nodes and complexity could introduce dynamics such as increased network congestion or efficiency improvements, depending on the model's design and management.
    \item \textbf{External Factors}: External factors such as network attacks, system outages, and regulatory changes can also impact the network's performance and were not considered in the simulation.
\end{enumerate}

\subsection{9.5 Conclusion}
Our SimPy-based simulation offers valuable insights into the throughput capabilities of the proposed dual-node Lightning Network payment system, suggesting a high degree of scalability and reliability. However, the real-world application of this system would likely experience variations in performance due to factors such as network congestion, variable transaction processing times, operational challenges, and external influences. Future work should focus on optimizing the network's resilience to these factors, possibly through advanced rebalancing strategies and enhanced node management, to ensure the network's robustness in diverse operational conditions. These simulation results provide a promising foundation, yet they should be viewed as an initial step towards a comprehensive understanding and continuous improvement of the network's performance.


\section{10. Regulatory Considerations}

The implementation of a dual central node system within the Lightning Network raises several regulatory considerations, particularly in relation to Money Transmission Laws (MTL). This section explores the potential classification of the central routing nodes as money transmitters, the implications of such a classification, and the corresponding regulatory requirements.

\subsection{10.1 Money Transmission Laws Overview}

Money Transmission Laws in the United States are formulated at the state level to safeguard consumers and mitigate the risks of financial crimes such as money laundering, fraud, and terrorist financing. At the federal level, the Bank Secrecy Act (BSA) [49] and regulations promulgated by the Financial Crimes Enforcement Network (FinCEN) delineate the obligations of money services businesses (MSBs), including money transmitters [50].

A money transmitter is generally defined as an entity engaged in the transfer of funds on behalf of the public by any means. This definition encompasses entities that provide payment processing services, facilitate payments between parties, and handle the transfer of value that substitutes for currency [51].

\subsection{10.2 Applicability to Central Routing Nodes}

Central routing nodes in a closed payment network on the Lightning Network may fall under the scope of money transmission regulations due to their role in facilitating the transfer of value between parties. As these nodes manage and route payments within the network, they exhibit characteristics similar to traditional money transmitters.

The dual central node approach introduces additional complexity, as both the active central routing node and the dormant node are integral to the network's payment processing capabilities. The dormant node, while typically inactive, serves as a backup that can assume the role of the central routing node, thus engaging in money transmission activities during switchover events.

\subsection{10.3 Regulatory Implications}

If central routing nodes are classified as money transmitters, the private payment facilitator operating the network must comply with a spectrum of regulatory requirements, including:

\begin{itemize}
    \item \textbf{Licensing}: Acquiring money transmitter licenses in each state where the network operates or serves clients, as mandated by state MTL statutes.
    \item \textbf{Compliance Programs}: Developing anti-money laundering (AML) and counter-terrorist financing (CTF) programs in accordance with the BSA and FinCEN guidelines, which include designating a compliance officer, conducting employee training, and performing regular audits [52].
    \item \textbf{Reporting and Recordkeeping}: Adhering to FinCEN's reporting requirements, such as filing Suspicious Activity Reports (SARs) and Currency Transaction Reports (CTRs), and maintaining comprehensive transaction records for a period of five years [53].
    \item \textbf{Consumer Protection}: Implementing measures to protect client funds, providing transparent terms of service, and ensuring fee structures are clear and reasonable, as per the Consumer Financial Protection Bureau (CFPB) regulations [54].
\end{itemize}

\subsection{10.4 Tax and Financial Management Complexities}

The use of Bitcoin as the medium of exchange in a closed payment network introduces complexities in tax and financial management for companies. These complexities arise from the evolving nature of cryptocurrency regulations and the lack of uniform guidance across jurisdictions.

The Internal Revenue Service (IRS) in the United States classifies virtual currencies such as Bitcoin as property for federal tax purposes, as detailed in IRS Notice 2014-21 [55]. Consequently, transactions involving Bitcoin are subject to tax implications similar to other forms of property, which include capital gains and losses reporting.

\begin{quote}
"The sale or other exchange of virtual currencies, or the use of virtual currencies to pay for goods or services, or holding virtual currencies as an investment, generally has tax consequences that could result in tax liability." - IRS Notice 2014-21 [55]
\end{quote}

Under U.S. tax law, specifically 26 U.S. Code § 6045, brokers and barter exchanges are required to report certain transactions to the IRS [56]. This includes the disposition of virtual currency that is held as a capital asset. The facilitator of a closed payment network may be considered a broker if they transfer Bitcoin on behalf of clients, thereby incurring reporting obligations.

\begin{quote}
"Every person doing business as a broker or barter exchange shall make a return according to the forms or regulations prescribed by the Secretary, setting forth the name and address of each customer..." - 26 U.S. Code § 6045 [56]
\end{quote}

The Financial Accounting Standards Board (FASB), which establishes accounting principles in the U.S., has not yet provided specific guidance on the accounting for cryptocurrencies. In the absence of such guidance, companies must navigate the accounting treatment of Bitcoin transactions using existing standards. This may involve determining whether Bitcoin is accounted for as cash, cash equivalents, financial instruments, intangible assets, or inventory under various applicable standards, including ASC 305, ASC 825, ASC 350, and ASC 330, respectively [57, 58, 59, 60].

For companies operating internationally, the tax treatment of Bitcoin can vary significantly by country. Cross-border transactions may trigger tax events in multiple jurisdictions, necessitating careful planning to ensure compliance with each country's tax code and avoiding double taxation.


\section{11. Decentralization Considerations}

The inception of Bitcoin and the subsequent development of the Lightning Network represent a paradigm shift towards decentralized financial systems. However, the implementation of a closed payment network, as discussed in this paper, introduces a centralized structure within the broader context of these inherently decentralized technologies.

It is important to recognize that while Bitcoin's protocol is decentralized, the actual deployment and operation of its infrastructure exhibit centralization tendencies. A significant portion of Bitcoin mining is controlled by large corporations due to the economies of scale in mining operations [61]. Additionally, nearly half of Lightning Network nodes are hosted on centralized cloud services such as Amazon Web Services and Google Cloud [62], which centralizes the physical infrastructure underpinning the network. Furthermore, the entry of funds into the Bitcoin ecosystem is often centralized, as individuals and businesses typically acquire Bitcoin through centralized exchanges requiring KYC or KYB, converting fiat currency into cryptocurrency unless it is directly mined or earned [63, 64].

\subsection{11.1 Centralization as a Gateway}

We posit that introducing businesses and customers to a centralized Lightning Network through a closed payment network can serve as an intermediary step towards a more decentralized future. The incentives for adoption are clear: the promise of rapid transaction speeds and lower fees compared to traditional payment methods. As businesses become accustomed to the efficiency and cost-effectiveness of such networks, they may be more inclined to hold a portion of their funds in Bitcoin. This gradual accumulation of Bitcoin holdings within the business sector could lead to a broader engagement with the decentralized aspects of the cryptocurrency ecosystem.

By facilitating the initial transition to a Bitcoin-based payment system within a controlled environment, businesses can build the necessary confidence and infrastructure to participate in the wider decentralized network. Over time, as the volume of transactions conducted over the Lightning Network grows, these businesses may become more integrated with the decentralized network, contributing liquidity and stability. This integration could potentially extend to the operation of independent nodes and participation in mining, further distributing the network's control and enhancing its decentralized nature.

In essence, a closed payment network could act as a gateway, lowering the barriers to entry for businesses and fostering a gradual shift from centralized financial systems to decentralized solutions. This approach aligns with the concept of bootstrapping, where centralized components are utilized to support the initial stages of a system that ultimately aims for decentralization. As businesses increasingly operate within the Bitcoin economy, their growing stake in the system could motivate a transition to more decentralized practices, contributing to the overall resilience and decentralization of the network 


\section{12. Conclusion and Future Research}

This paper has presented a dual-node configuration for optimizing closed payment networks on the Lightning Network. The proposed architecture employs a central routing node for transaction processing and a dormant node for rebalancing and backup, creating a system that is both efficient and resilient. This structure provides a viable solution for private payment facilitators seeking to establish a closed network environment that offers competitive advantages over traditional payment processors.

Throughout the paper, we have explored various aspects of the dual-node system, from network topology optimization to economic models and fee strategies. We have shown that the star graph topology, while necessitating two Lightning transactions for transfers between client nodes, offers a tradeoff that favors reduced operational complexity and cost efficiency. The manual channel management protocol ensures the closed network remains isolated from external transactions, preserving liquidity and upholding privacy.

The dual-node system's rebalancing mechanism, underpinned by circular rebalancing, enhances liquidity management without incurring excessive on-chain fees. The fault tolerance design, with the dormant node serving as a redundancy measure, provides the network with high availability, ensuring uninterrupted service.

The central routing node's economic model, with its potential for low or zero-fee transactions, positions the closed payment network as a strong competitor to established payment networks like Visa, Mastercard, and traditional bank wire transfers. The fast settlement times and low fees of the Lightning Network are unparalleled, offering significant benefits over existing payment methods, which often involve higher fees and longer processing times.

By leveraging the inherent advantages of the Lightning Network, private payment facilitators can provide a secure, private, and cost-effective payment system. The architecture outlined in this paper lays the groundwork for further research and development in this domain, with the potential to contribute significantly to the evolution and decentralization of payment systems in the digital age.

For future work, there are several topics to explore. These include:

\begin{itemize}
    \item Investigating alternative network topologies that may offer advantages in scalability, cost, and operational efficiency.
    \item Developing and evaluating additional rebalancing algorithms to optimize liquidity management.
    \item Exploring integration strategies with the broader public Lightning Network while preserving the closed network's integrity.
    \item Utilizing smart contract technology to further automate channel management and rebalancing processes.
    \item Conducting empirical studies to measure the performance impact of the dual-node architecture in real-world scenarios.
    \item Enhancing the security protocols to safeguard against potential threats specific to closed payment networks.
    \item Analyzing the use of cryptographic techniques to further enhance privacy.
\end{itemize}





\begin{thebibliography}{99}
\bibitem{CFR1010410} "Records to be made and retained by financial institutions," 31 CFR § 1010.410, Code of Federal Regulations.
\end{thebibliography}
\end{document}